\documentclass[aip,apl,english,twocolumn]{revtex4}

\usepackage{graphicx}
\usepackage{hyperref}
\usepackage{upgreek}
\usepackage{textcomp}
\usepackage{babel}

\begin{document}


\title{Guided Neuronal Growth on Arrays of Biofunctionalized GaAs/InGaAs
Semiconductor Microtubes} 



\author{Cornelius S. Bausch}
\email[]{cbausch@physnet.uni-hamburg.de}
\affiliation{Institut f\"{u}r Angewandte Physik, Universit\"{a}t Hamburg, 20355 Hamburg -- Germany}

\author{Aune Koitm\"{a}e}
\affiliation{Institut f\"{u}r Angewandte Physik, Universit\"{a}t Hamburg, 20355 Hamburg -- Germany}

\author{Eric Stava}
\affiliation{Institut f\"{u}r Angewandte Physik, Universit\"{a}t Hamburg, 20355 Hamburg -- Germany}
\affiliation{Department of Electrical Engineering, University of Wisconsin -- Madison, Madison 53706, Wisconsin -- United States}

\author{Amanda Price}
\affiliation{Department of Neuroscience, University of Wisconsin -- Madison, Madison 53706, Wisconsin -- United States.}

\author{Pedro J. Resto}
\author{Yu Huang}
\affiliation{Department of Biomedical Engineering, University of Wisconsin -- Madison, Madison 53706, Wisconsin -- United States}

\author{David Sonnenberg}
\author{Yuliya Stark}
\author{Christian Heyn}
\affiliation{Institut f\"{u}r Angewandte Physik, Universit\"{a}t Hamburg, 20355 Hamburg -- Germany}

\author{Justin C. Williams}
\affiliation{Department of Biomedical Engineering, University of Wisconsin -- Madison, Madison 53706, Wisconsin -- United States}

\author{Erik W. Dent}
\affiliation{Department of Neuroscience, University of Wisconsin -- Madison, Madison 53706, Wisconsin -- United States.}

\author{Robert H. Blick}
\affiliation{Institut f\"{u}r Angewandte Physik, Universit\"{a}t Hamburg, 20355 Hamburg -- Germany}
\affiliation{Department of Electrical Engineering, University of Wisconsin -- Madison, Madison 53706, Wisconsin -- United States}


\date{\today}

\begin{abstract}
We demonstrate embedded growth of cortical mouse neurons in dense arrays of semiconductor microtubes. The microtubes, fabricated from a strained GaAs/InGaAs heterostructure, guide axon growth through them and enable electrical and optical probing of propagating action potentials. The coaxial nature of the microtubes -- similar to myelin -- is expected to enhance the signal transduction along the axon. We present a technique of suppressing arsenic toxicity and prove the success of this technique by overgrowing neuronal mouse cells.
\end{abstract}


\maketitle 

Optical read-out of the interactions of protein arrays and cellular networks form the backbone of high-bandwidth parallel information processing techniques \citep{Schaeferling2006,Velasco-Garcia2009}. Hence, it is essential for bio-electronic circuitry –- such as microtubes discussed in this work –- to provide optically active materials, such as III/V-semiconductors. This enables optical tracing of propagating action potentials in cellular networks. Apart from the detection and stimulation of action potentials, sucessful
realization of neuronal guidance is the first step towards the designing
of neuronal networks in vitro. Guidance has previously been achieved
by using chemical guidance cues \citep{Greene2011,Staii2009} as well
as by exploiting the geometrical properties of the growth substrate \citep{Charrier2010,Lee2007}. It has been shown \citep{Yu2011a,Schulze2010}
that arrays of micrometer-sized silicon-based tubes can sucessfully
direct the outgrowth of neurons, where the neurites show a remarkable
attraction towards the tube orifices. 

So far, these microtubes have been fabricated from a strained bilayer
of Si/SiGe and Si/SiO$_{2}$, respectively, whereas we use a combination
of GaAs and InGaAs as a base material for the fabrication of arrays
of tubes. The usage of GaAs, an optical III-V semiconductor, for microtubes
offers a variety of advantages over Si: It exhibits a tunable, direct
bandgap, which makes it suitable for experiments with optogenetic
neurons. The growth of axons through optical bottle resonators \citep{Strelow2012}
or the exploitation of surface plasmons \citep{Rottler2011} could
spawn a new method of action potential detection.

Additionally, the electron velocity and mobility in GaAs are generally
higher than in Si, resulting in lower noise levels of electronic devices.
Together with other features such as its piezoelectric properties,
GaAs offers a whole new range of sensing mechanisms of action potentials
in neurons.

\begin{figure}
\begin{centering}
\includegraphics[width=.48\textwidth]{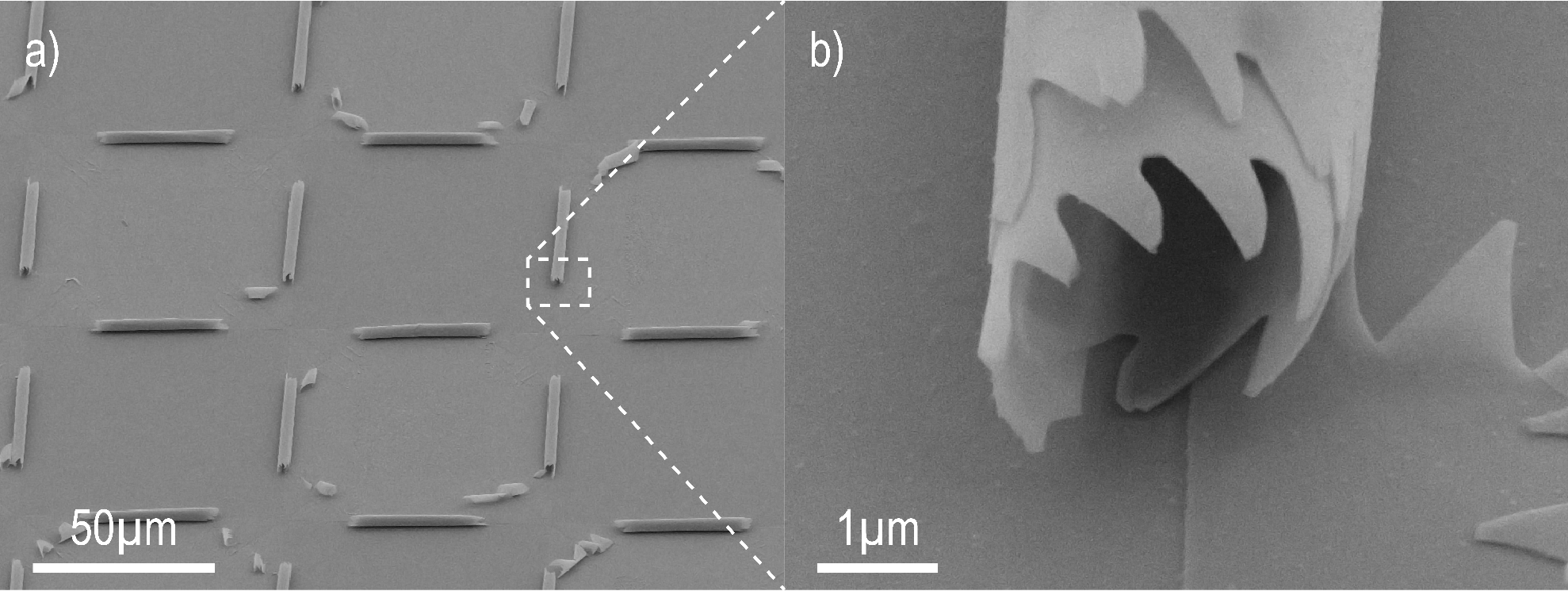}
\par\end{centering}

\caption{\label{fig:SEM-images} (a,b) SEM images of arrays of GaAs/InGaAs
heterostructure microtubes. The tubes are arranged in a square array
with tube lengths of $~$50\,$\upmu$m and distances between opposing
tube orifices of 40\,$\upmu$m. (b) The membrane at the end of the
tubes exhibits nonregular protrusions which are generated along the
crystal structure during the rolling-up process.}
\end{figure}

The basis for the fabrication of microtubes is a strained heterostructure
grown by molecular beam epitaxy on top of the $\left\langle 100\right\rangle $
surface of a GaAs substrate. It consists of a 40\,nm AlAs sacrifical
layer covered with layers of In$_{0.19}$Ga$_{0.81}$As and GaAs.
The layers are grown pseudomorphically, thus the different lattice
constants of In$_{0.19}$Ga$_{0.81}$As (5.73\,\AA) and GaAs (5.65\,\AA)
causes a strain in the top bilayer, which is released and leads to
the rolling up of the bilayer when the sacrificial layer is selectively
etched away using HF. The diameter of the microtubes is precisely
tunable via the choice of these different layer thicknesses. The diameter
of our tubes, determined by means of SEM imaging, is in the range
of about 2 to 5\,$\mu$m. In Fig. \ref{fig:SEM-images}(a,b), an
array of such microtubes is shown. The rate of successfully fabricated
microtubes on a sample was close to 95\% with about 5000 intact tubes,
thus it is possible to form highly complex neural networks on the
sample.

The arsenic component of GaAs, however, renders the material highly
toxic to cells \citep{Vigo2006,Frankel2009}. The 3-5\,nm thick,
native oxide layer consists of a nonuniform mix of As$_{2}$O$_{3}$,
Ga$_{2}$O$_{3}$ and elemental As \citep{Hou1997}. In aqueous solution,
these oxide layers dissolve and, due to their reformation with the
aid of oxygen in the water, a continuous etching process occurs. Pure,
cleaned GaAs corrodes in 140\,mM NaCl solution under incubating conditions
(37\textdegree{}C, 5\% CO$_{2}$) with etching rates of $\sim$200\,nm/day \citep{Kirchner2002}. Typical neuron cultures utilize Neural Basal
Medium with a concentration of 3\,g/l NaCl, which corresponds to
50\,mM NaCl. An efficient method of suppressing arsenic toxicity
is therefore imperative.

\begin{figure}
\begin{centering}
\includegraphics[width=.48\textwidth]{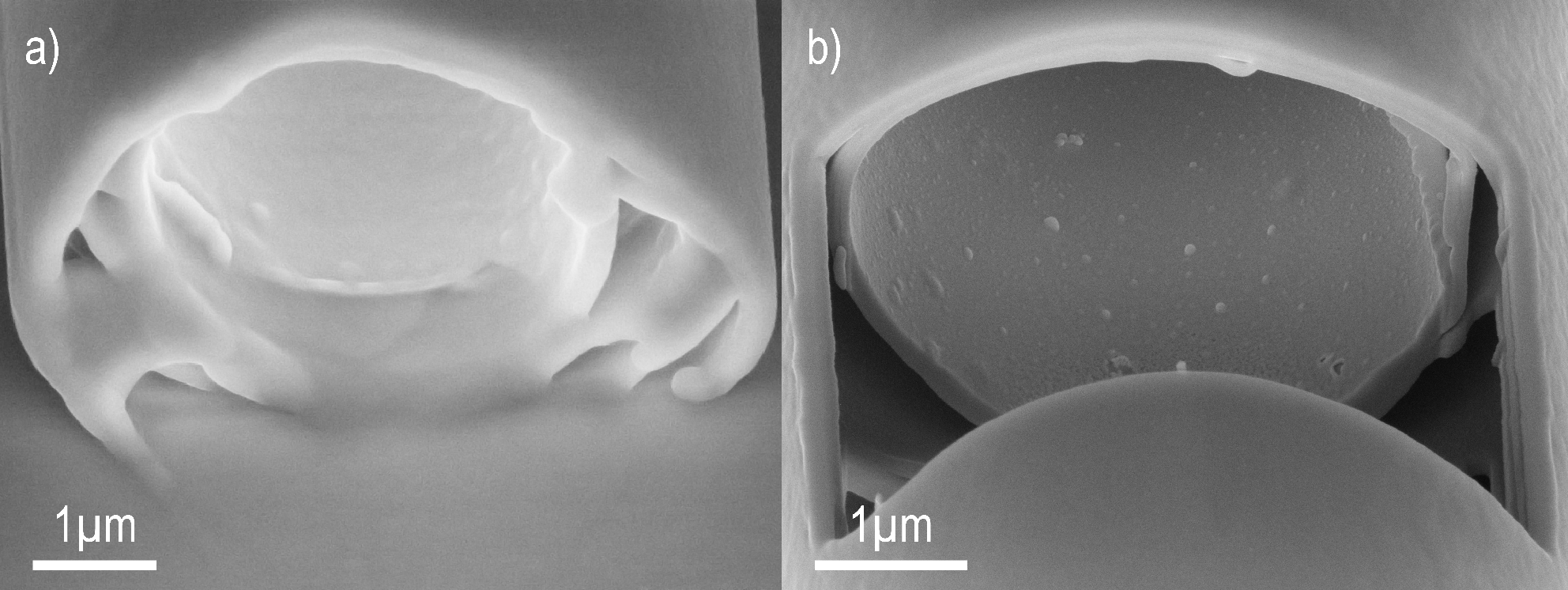}

\par\end{centering}

\caption{\label{fig:SEM-images-2}SEM images of the parylene-C coating of microtubes.
(a) Tube orifice coated with 160\,nm. (b) SEM image of the same parylene
coated 100\,$\mu$m long microtube cut open in the middle using a
FIB. The bubbles on the inner tube surface are regions of parylene
agglomeration thus proving the diffusion of parylene gas into the
tubes.}
\end{figure}

\begin{figure}[tb]
\begin{centering}
\includegraphics[width=.48\textwidth]{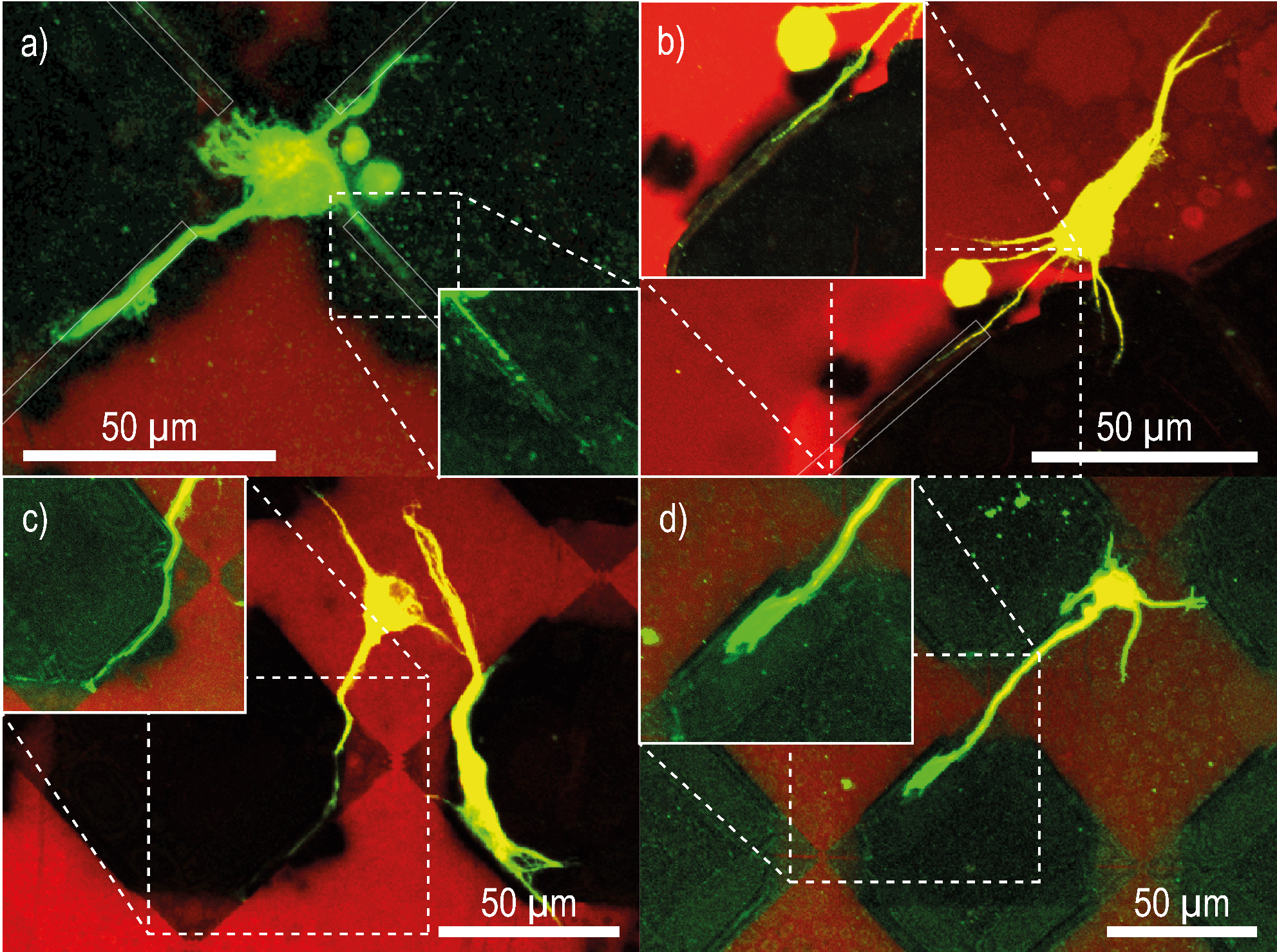}
\par\end{centering}

\caption{\label{fig:Neurons}Fluorescent microscopy images of cortical mouse
neurons grown for 1 day on a rectangular array of microtubes (marked
by the gray transparent lines) coated with 160\,nm of parylene. (a)
A soma attached to the substrate right in front of the orifices of
the four microtubes. The three extending neurites grew through the
tubes or along the tubes (see inset). (b) A neurite grows through
a microtube at the edge of the tube array. (c) After extending along
a tube, the neurite eventually finds a tube orifice to grow through.
(d) In some cases, the neurites do not outgrow through rather than
alongside the tubes.}
\end{figure}

Due to the release of arsenic compounds into the medium, the entirety
of the sample has to be protected from corrosion. This was achieved
by placing the samples onto semi-cured PDMS. After curing, the PDMS
with the wafer pieces was cut out to squares. The top part was covered
with parylene-C using the \textit{Specialty Coating Systems PDS2010}
parylene coater at a thickness of 160\,nm, as determined by means
of atomic force microscopy. The conformal CVD process ensures a diffusion
of the parylene monomer gas to the inner surface of the tubes, where
polymerization on the tube walls occurs. Fig. \ref{fig:SEM-images-2}(a)
shows the increase in the tube membrane thickness due to the parylene
coating (c.f. Fig. \ref{fig:SEM-images}(b)). Fig \ref{fig:SEM-images-2}(b)
proves the diffusion of parylene into the tubes and its polymerization
at the inner tube walls. 

The samples were sterilized under UV light for 30 min and coated in
0.1\,mg/ml Poly-D-Lysine (PDL) and, after one hour at room temperature,
they were rinsed three times with sterilized DI water. Dissected embryonic
primary cortical E15.5 mouse neurons (prepared following a previously
published protocol \citep{CultureProtocol}) were plated without
glia at a density of 5000 cells/cm$^{2}$ onto the samples together
with control substrates where the PDMS coating was omitted. After
incubation for 1 hour at 37\textdegree{}C and 5\% CO$_{2}$, the plating medium
was exchanged with serum-free medium. The cells were incubated for
up to 9 days where one third of the medium was exchanged every three
days.

\begin{figure}
\begin{centering}
\includegraphics[width=.48\textwidth]{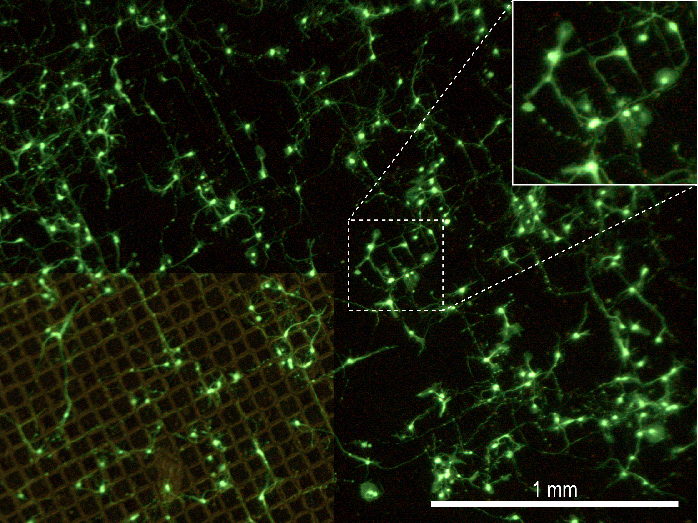}
\par\end{centering}

\caption{\label{fig:Neurons-2}Fluorescent microscopy image of a neuronal network
after growth for 9 days on an array of microtubes coated with 160\,nm
of parylene. The underlying square-shaped tube structure is resembled
in the outgrowth of the neuronal network (see inset on the lower left,
where the fluorescent image and the image of white light microscopy
are overlaid).}
\end{figure}

Despite the toxicity of GaAs, the neurons showed good viability after
culture for 1 day and 9 days, respectively. The viability was assessed
by fluorescent optical microscopy, where the live cells were stained
with the green-fluorescent calcein-AM and the dead cells were stained
with the red-fluorescent ethidium homodimer-1. The tubes attract the
outgrowth of neurites as shown in Fig. \ref{fig:Neurons}(a,b,c),
where neurons have grown neurites through microtubes. In some cases,
the neurites grew alongside the tubes rather than through them, as
Fig. \ref{fig:Neurons}(d) demonstrates. 

Growth on control substrates without the PDMS coating lead to the
death of all cells. This indicates the necessity of preventing the
corrosion of the sides and bottom of the growth substrate.

In contrast to recent results by others \citep{Yu2011a,Schulze2010},
we fabricated tubes with a smaller diameter, which is similar to the
size of a neuronal axon (2 to 3\,$\upmu$m) to ensure that only single
axons grew through the tube. The reduction of the tube diameter did
not have a negative effect on the attraction of neurites.

Figure \ref{fig:Neurons-2} shows a neuronal network on an array of
microtubes after culturing for 9 days. The long-term culture shows
a high density of living cells, and the overall appearance conforms
to the network of tubes (see inset in Figure \ref{fig:Neurons-2})
proving both the suppression of GaAs toxicity and cell guidance through
microtubes.

In this letter, we demonstrated that tube structures made of a GaAs/InGaAs
heterostructure can be used to guide the outgrowth of neuronal networks
where, as a first step, the highly toxic effects of arsenic compounds
were successfully suppressed using a combination of parylene and PDMS
coating. For future work, it is necessary to compare the viability
of neurons on GaAs-based microtubes to those based on Si/SiGe via
statistical analysis. Also, for a better outgrowth of the neural network,
the cell attachment has to be confined to the area in front of the
orifices of the microtubes, which could be achieved by PDL patterning.

\bibliography{Bausch_MANUSCRIPTarxiv}

\end{document}